\documentclass[prl,twocolumn,showpacs,superscriptaddress]{revtex4}

\usepackage[dvips]{graphicx}
\usepackage{color}
\usepackage{latexsym}
\usepackage{amsmath}
\usepackage{amsthm}
\usepackage{amssymb}
\usepackage{amsfonts}

\begin{document}
\title{ Fermi liquid behavior and Luttinger's theorem close to a diverging scattering length}
\author{Sergio~Gaudio}
\affiliation{Universita' degli Studi di Roma, La Sapienza and ISC-CNR, Roma, 00185, Italy}
\affiliation{Department of Physics, Boston College, Chestnut Hill, MA, 02467}
\author{Jason Jackiewicz}
\affiliation{Max Planck Institute for Solar System Research, Katlenburg-Lindau, Germany}
\author{Kevin S. Bedell}
\affiliation{Department of Physics, Boston College, Chestnut Hill, MA, 02467}
\begin{abstract}
Based on the results obtained in a previous paper \cite{GJB}, we derive the thermodynamic properties of a Fermi gas, deep into the quantum degenerate regime.  We show that, if Luttinger's theorem holds, a first order phase transition occurs in the normal phase as a function of the interaction strength, U. We also show that a volume change occurs at finite temperatures from the BEC to the BCS side of a diverging s-wave scattering length, in the normal phase. The transition has an end point above the BCS critical temperature. Also we show that a paramagnetic system in equilibrium, close to the divergence of the scattering length, on the negative side, screens out any applied magnetic field. 
\end{abstract}

\maketitle
\begin{figure}[!b]
 \includegraphics[width=3.5in] {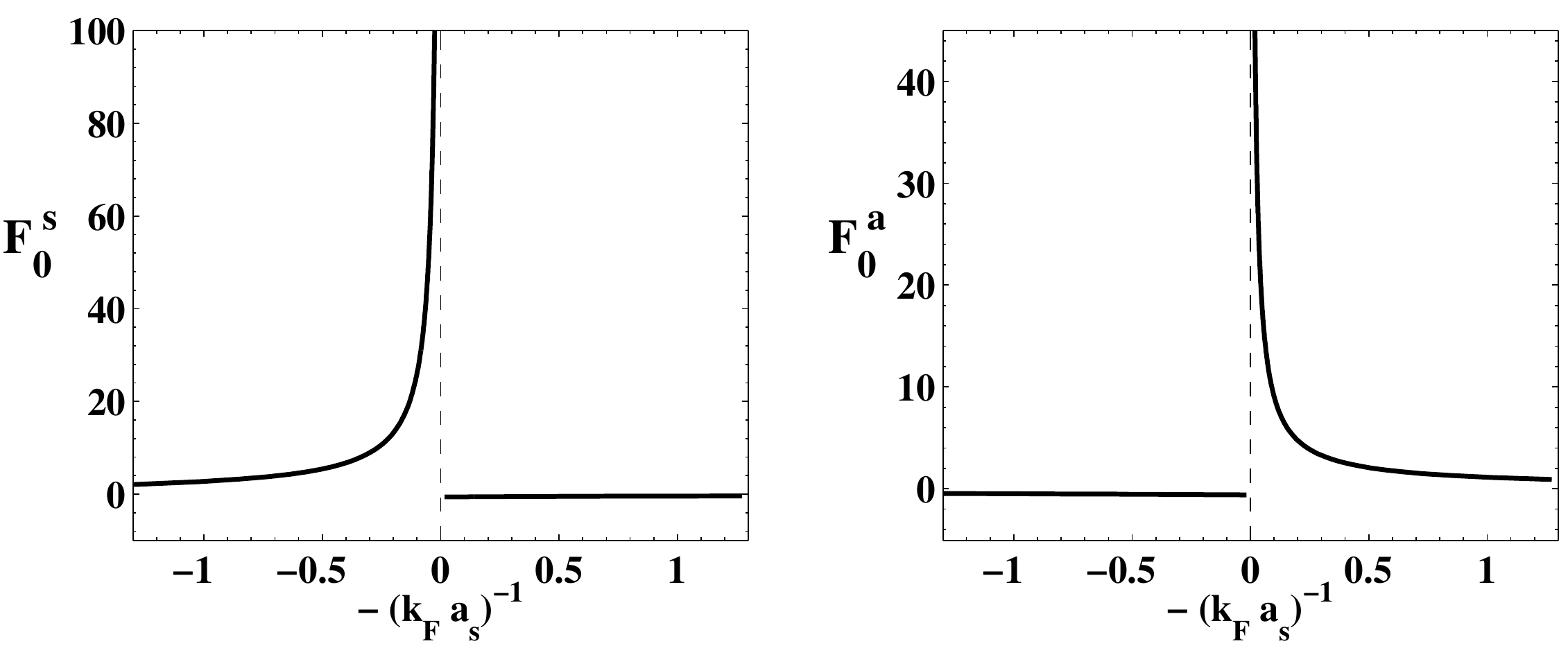}
	\caption{\label{4fs0}.  The first two Fermi liquid parameters, $F^{s,a}_0$. The parameters enter directly in the definition  compressibility and spin susceptibility as seen in text. On the BEC (BCS) side, $F^{a(s)}_0 \sim -.6$.}
\end{figure}
\begin{figure}[!b]
 \includegraphics[width=3.2in] {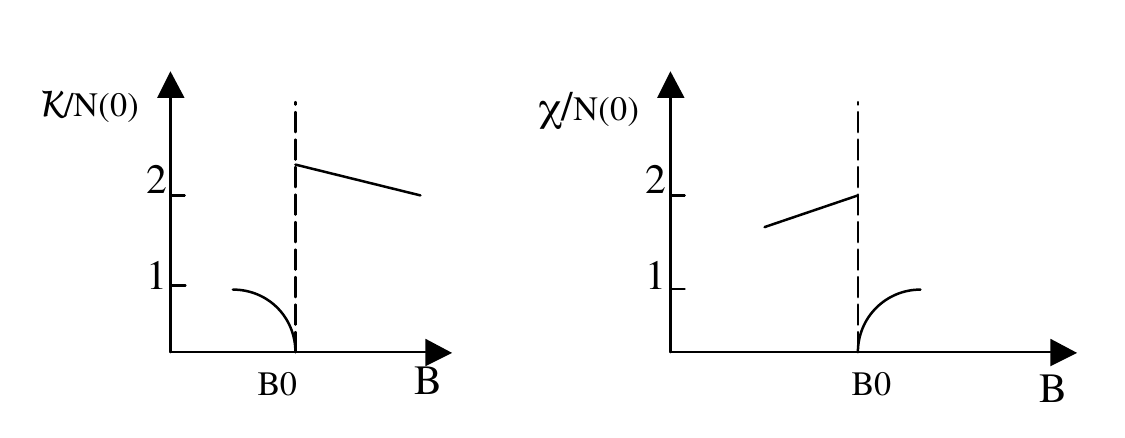}
	\caption{\label{k_chi}. A schematic view of the compressibility and spin susceptibility in the vicinity of the divergence.}
\end{figure}

Fermi liquid theory is characterized by an infinite number of unknown parameters, the Landau (or Fermi liquid) Parameters (LP) which are hard  to derive  from an underlying microscopic hamiltonian \cite{nozieres}.
This is even more pronounced in  such strongly correlated fermionic systems  as nuclear matter,  neutron stars, $^3$He, overdoped high T$_c$ superconductors and QCD just to name a few. In this context, an important contribution could come from the cold atom physics community. Feshbach resonances constitute a unique opportunity to study this kind of systems in the strongly interacting region, because of the easy experimental accessibility.  Still, two component Fermi gases, have mostly been studied in terms of the superfluid phase and much less attention has been given to the normal one. The possibilities offered, though,  are finally being explored also in the normal phase, particularly to the MIT group \cite{schunck}, to which our calculations could be related to some extent.   
Theoretically, the strongly interacting limit is studied in terms of mean field theory or in terms of the Random Phase Approximation (RPA), which, as we showed\cite{GJB}, gives the wrong physics close to resonance. As a consequence, these methods provide, at best, a qualitative answer for weakly and intermediate interacting regimes.\\
\indent In Ref \cite{GJB}, we showed how to derive from a microscopic theory the parameters of a Fermi liquid \cite{comment} in the full range of interactions, from weak to strong coupling. Our calculations are exact  in that they satisfy the Landau and Bedell-Ainsworth sum rules \cite{B_sumrule} in the low energy regime ($T_c < T\ll T_F$, where $T_c$ is the temperature of the BCS transition and $T_F$ is the Fermi temperature)\cite{commbec} and non-perturbative, since they do not depend on the existence of any small parameter.\\
\indent Before going to the polarized case, which we will discuss somewhere else, we will explore the consequences of a Fermi liquid theory close to a diverging scattering length for the case of equal population systems, since many results might be equivalent.\\
\indent In this Letter, we show that the compressibility, the spin susceptibility and the specific heat have a jump when the divergent s-wave scattering lenght changes sign and that this jump might be related to a first order phase transition since it gives rise to a non-zero latent heat.  If Luttinger's theorem \cite{FS}  holds, we show that the presence of a divergent s-wave scattering length implies a change in the volume on the two sides of the resonance \cite{notation}. We looked for the existence of a end point in the transition and found that this temperature is above the superfluid critical temperature. Thus,  the absence of the transition would signify that this theorem does not hold when the interaction strength changes sign. Perhaps, the most relevant consequence of our calculations is the spontaneous creation of pairing on the negative side of the strongly interacting regime, which for the HTSC would signify the appearance of a pseudogap-like region, below the temperature, T$^* \lesssim .35$ T$_F$.  The pseudogap, thus, would result as a consequence of the nature of the Fermi liquid state rather than from fluctuations (thermal/quantum) of the superfluid state and the breaking of Fermi liquid theory. Also, since the spin susceptibility vanishes close to the infinite negative value of the scattering length, a small magnetic field , i.e. as long as  $\mathbf{B}_{eff}=(\chi/N(0))  \mathbf{B}$, where $\mathbf{B} (\mathbf{B}_{eff})$ is the (effective) field  , $\chi$ is the spin susceptibility and $N(0)$ is the density of states (DOS), applied to the gas in equilibrium cannot polarize the system.\\
\indent  Following \cite{GJB}, our calculations for the LP are shown in Fig.\ref{4fs0}.
It is interesting to look at the behavior of the  parameters. 
In the limit $a_s \rightarrow + \infty$, the spin fluctuations are dominant ($F^s_0 \rightarrow \infty$), whereas, in the limit $a_s \rightarrow -\infty$, their role is taken over by the density fluctuation ($F^a_0 \rightarrow \infty$). This has immediate consequences on the respective thermodynamic quantities.  From \cite{FLT}, the compressibility and the spin susceptibility are given by  
\begin{eqnarray}
\mathcal{K} = \frac{N(0)}{1+F^s_0}, \\
 \chi  =  \frac{N(0)}{1+F^a_0}.
\end{eqnarray}
Therefore, these quantities are discontinuous as we go through the divergence of $a_s$. In Fig.\ref{k_chi} we sketch this behavior.
 \\
\indent The fact that some Fermi liquid parameters diverge may be a little surprising at first. We can understand this, by looking at the Bedell-Ainsworth sum rule, 

\begin{equation}
\label{B-sum_rule}
\frac{4}{\pi}\frac{m^*}{m}k_F a_s = \sum_l\left(F^s_l -F_l^a \left(1+2\frac{F^a_l/(2l+1)}{1+F^a_l/(2l+1)}\right)\right),
\end{equation}

\noindent where $F^{s (a)}_l$ the superscript indicates the symmetric (antisymmetric) spin-rotational component and the subscript denotes the Legendre component.  From this, it follows that if the scattering length diverges, the RHS has to diverge. The sum rule is usually\cite{Pom} exhausted by the first few momenta of the Landau expansion.  As a consequence, at least one of $F^{s,a}_{l}$ has to diverge. From our model,  the diverging LP is $F^{s}_0$ upon approaching $a_s = +\infty $  and $F^a_0$ upon approaching $a_s = -
\infty$ .\\
\indent Another of the LP that we are able to calculate is $F^s_1$, which in a Galilean invariant system is related to 
 the effective mass, $m^*$,  through  \cite{FLT}
 
 \begin{equation}
 m^*/m = 1+ \frac{F^1_s}{3}.
 \end{equation}

\begin{figure}[!t]
 \includegraphics[width=2.8in] {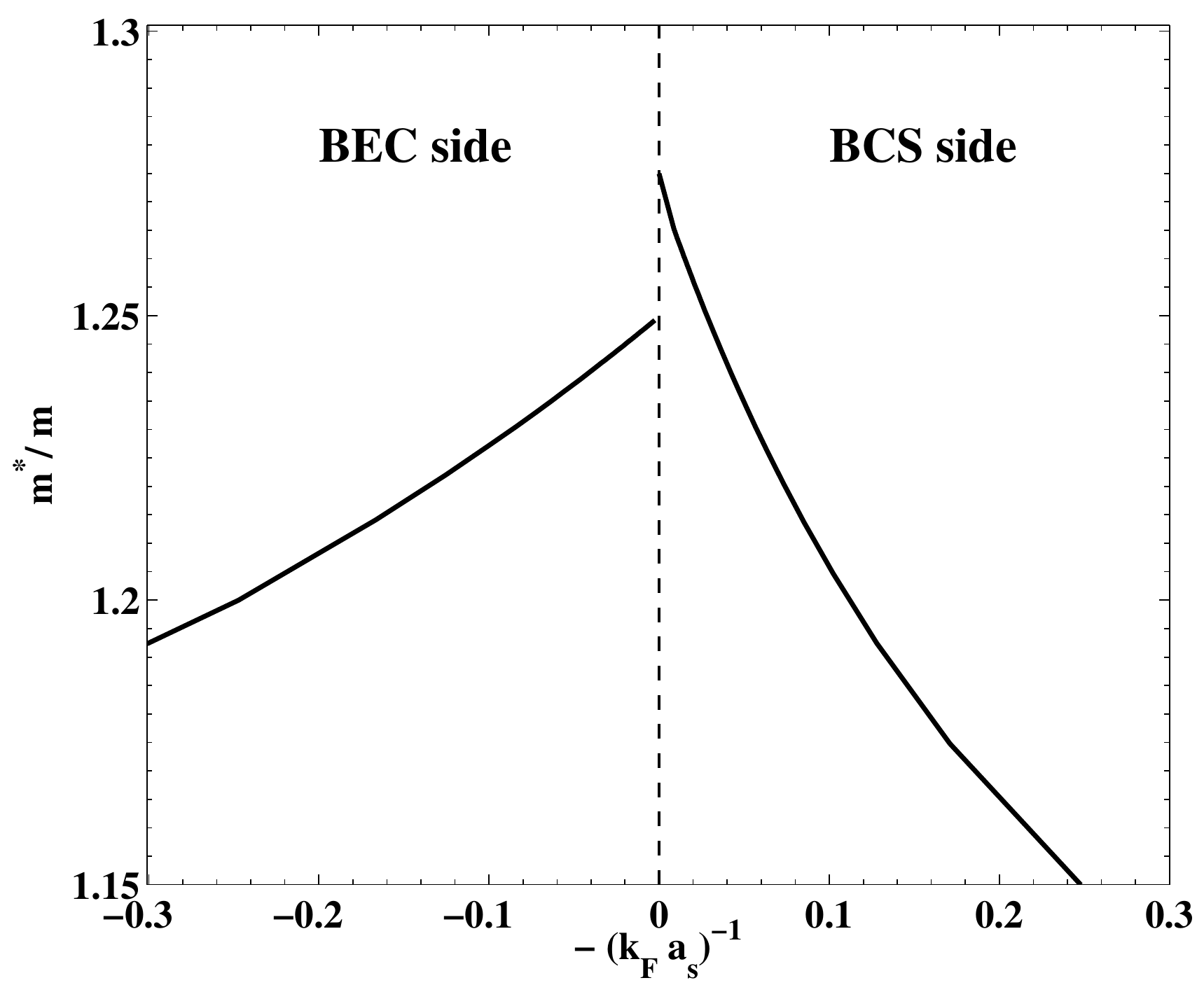}
	\caption{\label{emass_chi}. Effective mass, $m^*$, close to $a_s = \pm \infty$ .  The small jump indicates a "weak" first order phase transition.}
\end{figure}

\noindent The change in the effective mass is shown in Fig.\ref{emass_chi}.
 The jump around $|U|^{-1} = 0$  ($U =(4\pi\hbar^2/m) a_s$)  in $m^*$ leads to a small discontinuity in the specific heat.
From Fig.\ref{4fs0} and Fig.\ref{emass_chi} it is clear that the many body effects become important near  $1/U = 0$ and decrease rapidly far away from the resonance. \\
\indent In order to identify the transition, we calculate the latent heat . This is defined as

\begin{equation}
\label{LHdef}
L = T \Delta S = \frac{p_F}{3} \frac{1}{\hbar^3}k^2_B T^2\left[m^*(-\infty)-m^*(+\infty)\right],
\end{equation}

\noindent where $\Delta S$  is the variation of the entropy. Here $L = L(a_s)$ where $a_s$ is the s-wave scattering length. The latent heat is calculated across the divergence of $a_s$ and is different from zero at $T \neq 0$. A latent heat stored means the occurrence of a first order phase transition. 
This transition would be similar to a first order liquid-vapour transition  in water, since there is no change in the translational symmetry.\\
\indent Let's look now at the Clausius-Clayperon equation, that is 

\begin{equation}
\frac{d P}{d T} = \frac{\Delta S}{\Delta V},
\end{equation}

\noindent where 
$P$ is the pressure, $V$ is the volume and $S = \gamma T$ is the entropy of a Fermi liquid at low temperature. The volume has a temperature dependence that in leading orders can be written as

\begin{equation}
\label{VT}
V = V_0 -\alpha \frac{T^2}{2},
\end{equation}

\noindent where
  
\begin{equation}
\alpha = \frac{1}{2}  \left[\frac{\partial (\gamma V)}{\partial P}\right]_{T=0, N},
\end{equation}

\noindent and $\gamma = \pi^2 N(0)/3$.
\noindent A little algebra gives

\begin{equation}
\label{alpha2}
 \alpha = \frac{V_0}{2}\left(\gamma\mathcal{K}\right)+ \frac{V_0}{2}\frac{\partial \gamma}{\partial P}.
\end{equation}
\begin{figure}[!t]
 \includegraphics[width=2.8in] {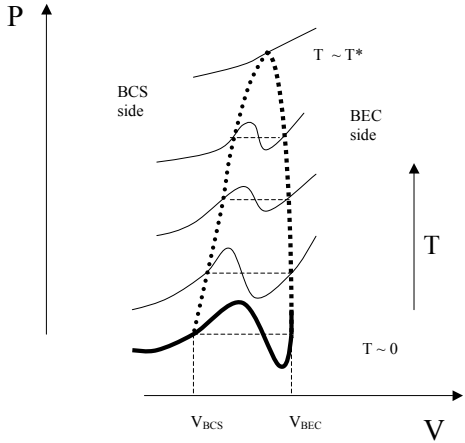}
	\caption{\label{pic1} Pressure-Volume. The qualitative isotherms depict the situation at $|U|^{-1} = 
 0 $ for increasing temperatures from bottom to top figure.}
\end{figure}

\noindent From Fig.\ref{4fs0}, then the compressibility is zero, at resonance, on the BEC side and finite on the BCS side and therefore we expect an infinite slope on the BEC side given by the Clausius-Clayperon equation and a finite slope on the BCS side, as seen in Fig.\ref{pic1}.
This gives a physical intuition of what happens close to the resonance on the two different sides. When  $a_s = +\infty $, $\alpha$ is basically zero for $T \rightarrow 0$, since  $\gamma$ close to resonance, on either side, is slowly varying.  For $a_s = -\infty$, the compressibility is finite, therefore this implies a jump in the curvature of the P-V diagram. From Eq.(\ref{VT}), the volume has non-zero contribution from the coefficient of expansion at $T \neq 0$ and appears to be discontinuous on the two sides of the resonance. This is shown, qualitatively, in Fig.\ref{pic1}. 
In a liquid-gas transition in water, there exists an end point, which is called the critical point, where the latent heat vanishes. 
In order to estimate it in our system, we need to study the correction to the entropy, $\delta S$, as a function of $T$, beyond the leading order, this is given by \cite{FLT}
\begin{equation}
\label{DS}
\delta S = -\frac{1}{20}\pi^4 n k_B B \left(\frac{T}{T_F}\right)^3 \log \left(\frac{T}{T^*}\right),
\end{equation}
where
\begin{equation}
\label{B}
B = -\frac{1}{2}\left[ (A^s_0)^2\left( 1- \frac{\pi^2}{12}A^s_0\right)+3 (A^a_0)^2\left( 1- \frac{\pi^2}{12}A^a_0\right)\right],
\end{equation}
and $T^*$ is related to some cut-off energy. In Eq.(\ref{B}), $A^{s,a}_0$ are the symmetric and antisymmetric components of the scattering amplitudes for the Legendre component $l =0$.
 We assume $T^*$ to be the energy for which we obtain the correct transition temperature to the BCS superfluid phase, $T_c \sim .2 T_F$. Our cut-off is, thus, $T^* \sim .35 T_F$\cite{qstar}.  Therefore, the critical point of the first order phase transition is above the superfluid transition temperature. 
 In Fig.\ref{pic1}, we show schematically the phase diagram for different isotherms. For $T\neq 0$, we are not able to predict how the system behaves, but it is reasonable to assume that as one compresses the system slowly, one should observe a mixture of the two phases in analogy with the liquid-vapour transition. As pointed out above, the transition and the effects we discuss are, for the most part, below or at temperature comparable to the superfluid transition one and it would seem hard to observe them, for example, in cold atom systems. A favorable experimental situation, though, is already available. In polarized systems \cite{polarized}, it has been shown that it is possible to suppress the superfluid phase. The resulting system, although not corresponding exactly to the situation at hand here, is a polarized Fermi liquid at  zero temperature. \\
\indent Up to now, we have implicitly assumed that $k_F$ is the same on the two sides of the divergence, that is, we have assumed the validity of the Luttinger's theorem as we go across the resonance . If this is true, then

\begin{equation}
\label{}
\frac{N_{BEC}}{V_{BEC}}= \frac{N_{BCS}}{V_{BCS}}.
\end{equation}

Therefore, a change in the volume, due to the transition we predict, implies a change in the number of particles on the two sides of the resonance.
In cold atoms traps, though, the system has the possibility to convert these unpaired fermions into composite  bosons or paired fermions, either in the closed channel or in the open channel.  It is known \cite{hulet05} that in $^6$Li some molecules are still present at resonance, in the closed channel, whereas, for equal (unequal) populations systems, paring has also been observed in the absence of a superfluid phase \cite{grimm} (\cite{schunck}). Therefore, the percentage of particles lost in these channels must be equal to the volume change of the Fermi liquid, in order to have the same Fermi momentum.  Since only a small part of the Fermi surface is involved, the occupied volume change might not be detectable.\\
\indent The first order transition we have described, is not the only possibility for the system. If Luttinger's theorem does not hold, the system could still avoid the more dramatic first order transition provided

\begin{equation}
\label{ }
k^{BEC}_F\; m^*_{BEC} = k^{BCS}_F\; m^*_{BCS}. 
\end{equation}

Since $m^*_{BCS} > m^*_{BEC}$, it must be $k^{BCS}_F <  k^{BEC}_F$. This implies that the density of  fermions in the system has to be somewhat smaller on the BCS side. Keeping the volume constant, this implies that
\begin{equation}
\label{ }
\left(\frac{m^*_{BCS}}{m^*_{BEC}}\right)^3 =\frac{N_{BEC}}{N_{BCS}}.
\end{equation}
 
\noindent Since $\left(\frac{m^*_{BCS}}{m^*_{BEC}}\right)^3\sim 2\%$, there is a small change in the number of fermions on the two sides, or better in the density, which means a small change in the Fermi surface. In this case, the latent heat would be zero. In both cases,  it is quite relevant to emphasize that in the absence of a closed channel, as in HTSC for example, the system can create pairs in the stronlgy interacting region, which will (only) appear as a pseudogap just below our T$^*$, but have nothing to do with fluctutions of the superfluid state, as it seems the case in \cite{schunck} .
Also, by looking at the properties of the system close to the resonance, we would implicitly test the Luttinger's theorem. 

It is important to emphasize that whether  Luttinger's theorem does or does not  hold, still the system requires that the two Fermi liquids be composed of two different number of fermions. Thus, if there is a mass difference on the two sides of the resonance, only a first order transition is the signature of the validity of Luttinger's theorem.

In recent experiments \cite{polarized}, it was found that a s-wave BCS state existed for large spin population imbalances, without, up to now,  any clear evidence for exotic superfluid phases. In such polarized system, one might expect to be beyond the Clogston-Chandransekhar limit \cite{clogston}. In fact, naively, based on our calculation, we would argue that the system should not be in an s-wave superfluid state at all.
To see this, we note that in Ref.\cite{hari}, it was pointed out that for small polarizations, $m = (n^\uparrow -n^\downarrow) \ll n$, where $n^{\sigma}$ is the number of quasiparticles of spin $\sigma$,  in quasi-equilibrium systems, the quasi-particles feel an effective internal field given by  $(F^a_0 m)/ N(0)$. Clearly this field, for any small $m$, would be larger than $\Delta$, the superfluid gap, close to resonance, since $F^a_0$ diverges. This would seem counter to recent experiments\cite{polarized}, where an s-wave BCS state was found close to resonance for a broad range of polarizations. In those experiments, though, the system phase separates in two regions, one with high polarizations and no superfluidity, and one with s-wave superfluidity and no polarization. Based on the argument we gave above, this is indeed the only way to achieve an s-wave superfluid in a polarized system near resonance. Given the fact that this is a superfluid with a short coherence length, this would allow the system to remain in the superfluid state for large population imbalances.\\
\indent In conclusion, a Fermi liquid close to a diverging scattering length offers some unexpected behavior. By deriving microscopically the Fermi liquid parameters, we observed a  jump across the divergence of $a_s$ in the compressibility, spin susceptibility, and specific heat . We showed that this  transition corresponds to a first order transition, since the latent heat is finite and non zero at resonance. We showed that this transition is related to the  validity of Luttinger's theorem, thus we provide implicitly a tool for testing it.  
Most importantly, we showed that whether the latter theorem does or does not hold, Fermi liquid theory imposes a "pseudogap-like" region below T$^*$ in the crossover region which is strictly a consequence of the thermodynamics of the Fermi liquid and \emph{not} due to fluctuations of the superfluid state. We also provide a plausible physical argument for the observation of the s-wave superlfuidity close to resonance and over a broad range of population imbalance. We are currently exploring the consequences of this theory for large polarizations in more detail. 


\begin{thebibliography}{99}
\bibitem{GJB}
S. Gaudio, J. Jackiewicz, K. Bedell, cond-mat/0505309
\bibitem{nozieres}
P. Nozieres, \emph{Theory of interacting fermions systems}, 1997 Addisons and Wesley Books;
\bibitem{schunck}
C. Schunck \emph{et al} cond-mat/0702066 
\bibitem{comment}
See \cite{FLT} for a review on Fermi liquid theory.
\bibitem{B_sumrule}
K. S. Bedell and T. L. Ainsworth, Phys. Lett. \textbf{102}A, 49 (1984);
T. L. Ainsworth, K. S. Bedell, Phys Rev. B, \textbf{35}, 8425 (1987).
\bibitem{commbec}
In the atomic Fermi gases, it is possible to have a Fermi liquid also on the BEC side, just very close to resonance for $T > E_b$ where $E_b$ is the molecular  binding energies. From mean field theory, $E_b = 0$ for $a_s\rightarrow \infty$.
\bibitem{FS}
The free gas Fermi Surface volume (FS) is preserved when the interaction is turned on, see J.M. Luttinger, Phys. Rev. \textbf{119}, 1153 (1960) .
\bibitem{notation}
For clarity, we indicate with the term "resonance", the point in which $|a_s| = \infty$, which in atomic systems is realized through Feshbach resonances (the way an infinite scattering length is achieved is not important for our work) and with BEC (BCS) side, the region where the scattering length is positive(negative), as it is customary in cold atom physics
\bibitem{FLT}
G. Baym and C. Pethick, \emph{Landau-Fermi liquid theory}, 1991, Wiley and Sons;
P. Nozieres and D. Pines, \emph{The Theory of Quantum liquids}, 1999, Perseus Books.
\bibitem{Pom}
This is true only in the absence of Pomeranchuck instabilities, I. Ia. Pomeranchuk, Sov. Phys. JETP \textbf{8}, 361(1958).
\bibitem{Babu} 
S. Babu and G. E. Brown, Ann. Phys., \textbf{78}, 1, (1973).
\bibitem{ABBQ}
T.L. Ainsworth \emph{et al.} J. Low Temp. Phys., \textbf{50}, 319 (1983).
\bibitem{pethick}
C.J. Pethick and H. Smith, \emph{Bose-Einstein Condensation in Dilute Gases}, 2002, Cambridge University Press.
\bibitem{hulet05}
G.B. Partridge \emph{et al.}, cond-mat/0505353.
\bibitem{grimm}
C. Chin \emph{et al.}, cond-mat/0405632
\bibitem{AGD}
A.A. Abrikosov, L.P. Gorkov, and I.E. Dzyaloshinski, \emph{Methods of Quantum Field Theory In Statistical Physics}, 1963, Dover.
\bibitem{Landau}
L. D. Landau,  E. M Lifshitz, \emph{Fisica Statistica}, Editori Riuniti Edizioni Mir,1981 
\bibitem{qstar}
The momentum of the collective modes corresponding to this critical temperatures is of the order $\bar q^* \sim .1 k_F$. Thus, $\bar q^* \ll k_F$
\bibitem{polarized}
G. B. Partridge, et al., Science \textbf{311}, 503 (2006);
M. W. Zwierlein, et al., Science \textbf{311}, 492 (2006)
%
\bibitem{hari}
K. S. Bedell, H. P. Dahal, Phys. Rev. Lett., \textbf{97}, 047204 (2006).
\bibitem{clogston}
A. M. Clogston, Phys. Rev. Lett. \textbf{9}, 266 (1962); B.S. Chandransekhar, App.Phys. Lett. \textbf{1}, 7 (1962)
\end{thebibliography}
 \end{document}